\def\year{2022}\relax
\documentclass[letterpaper]{article} % DO NOT CHANGE THIS
\usepackage{aaai22}  % DO NOT CHANGE THIS
\usepackage{times}  % DO NOT CHANGE THIS
\usepackage{helvet}  % DO NOT CHANGE THIS
\usepackage{courier}  % DO NOT CHANGE THIS
\usepackage[hyphens]{url}  % DO NOT CHANGE THIS
\usepackage{graphicx} % DO NOT CHANGE THIS
\usepackage{natbib}  % DO NOT CHANGE THIS AND DO NOT ADD ANY OPTIONS TO IT
\usepackage{caption} % DO NOT CHANGE THIS AND DO NOT ADD ANY OPTIONS TO IT
\DeclareCaptionStyle{ruled}{labelfont=normalfont,labelsep=colon,strut=off} % DO NOT CHANGE THIS
\usepackage{algorithm}
\usepackage{algorithmic}

\usepackage{newfloat}
\usepackage{listings}
\floatstyle{ruled}
\newfloat{listing}{tb}{lst}{}
\floatname{listing}{Listing}
\title{An Experience Report of Executive-Level Artificial Intelligence Education \\in the United Arab Emirates}
\author{
    David Johnson\textsuperscript{\rm 1},
    Mohammad Alsharid\textsuperscript{\rm 2},
    Rasheed El-Bouri\textsuperscript{\rm 2},
    Nigel Mehdi\textsuperscript{\rm 3},
    Farah Shamout\textsuperscript{\rm 4},    
    Alexandre Szenicer\textsuperscript{\rm 5},
    David Toman\textsuperscript{\rm 6},
    Saqr Binghalib\textsuperscript{\rm 7}
}
\affiliations {
    \textsuperscript{\rm 1}Department of Informatics and Media, Uppsala University, Box 513, 751 20, Uppsala, Sweden\\
    \textsuperscript{\rm 2}Department of Engineering Science, University of Oxford, Oxford, OX3 7DQ, United Kingdom,\\
    \textsuperscript{\rm 3}Department of Computer Science, University of Oxford, Oxford, OX1 3QD, United Kingdom,\\
    \textsuperscript{\rm 4}Engineering Division, New York University Abu Dhabi, 129188, United Arab Emirates\\
    \textsuperscript{\rm 5}Department of Earth Sciences, University of Oxford, Oxford, OX1 3AN, United Kingdom\\
    \textsuperscript{\rm 6}Linacre College, University of Oxford, Oxford, OX1 3JA, United Kingdom,\\
    \textsuperscript{\rm 7}Artificial Intelligence Office, Emirates Towers, Dubai, United Arab Emirates\\
    david.johnson@im.uu.se,
    \{mohammad.alsharid, rasheed.el-bouri\}@eng.ox.ac.uk,
    nigel.mehdi@cs.ox.ac.uk,\\
    fs999@nyu.edu,
    \{alexandre.szenicer, david.toman\}@linacre.ox.ac.uk,
    saqr.binghalib@ai.gov.ae
}

\begin{document}

\maketitle

\begin{abstract}
\begin{quote}
Teaching artificial intelligence (AI) is challenging. It is a fast moving field and therefore difficult to keep people updated with the state-of-the-art. Educational offerings for students are ever increasing, beyond university degree programs where AI education traditionally lay. In this paper, we present an experience report of teaching an AI course to business executives in the United Arab Emirates (UAE). Rather than focusing only on theoretical and technical aspects, we developed a course that teaches AI with a view to enabling students to understand how to incorporate it into existing business processes. We present an overview of our course, curriculum and teaching methods, and we discuss our reflections on teaching adult learners, and to students in the UAE.
\end{quote}
\end{abstract}

\section{Introduction}

\noindent Educational offerings on artificial intelligence (AI) have expanded hugely in the last 10 years. This has, in part, been driven by AI moving into the mainstream of public discourse in a world now dominated by data. Gaining an understanding of AI is now broadly recognized as being important at all educational levels. 

Learning about AI in formal education has, for the most, part been confined to university-level courses, primarily as a sub-discipline within computer science, and is taught now more broadly and applied in domains such as business, medicine, biology, engineering, and even the social sciences. In fact, according to Stanford University's AI Index Report 2021, a survey conducted in 2020 indicates that the world’s top universities have increased their AI education offerings in the past 4 years \cite{zhang_ai_2021}. Today, AI education is further being considered at the full range of educational levels, from pre-K and kindergarten, through K-12 and university, and to adults engaged in continuing education and continuing professional development programs. Settings outside of traditional university degree courses are being used to teach AI to younger audiences \cite{williams_popbots_2019}, as well as to adults.

\begin{figure}[t]
\includegraphics[width=\columnwidth]{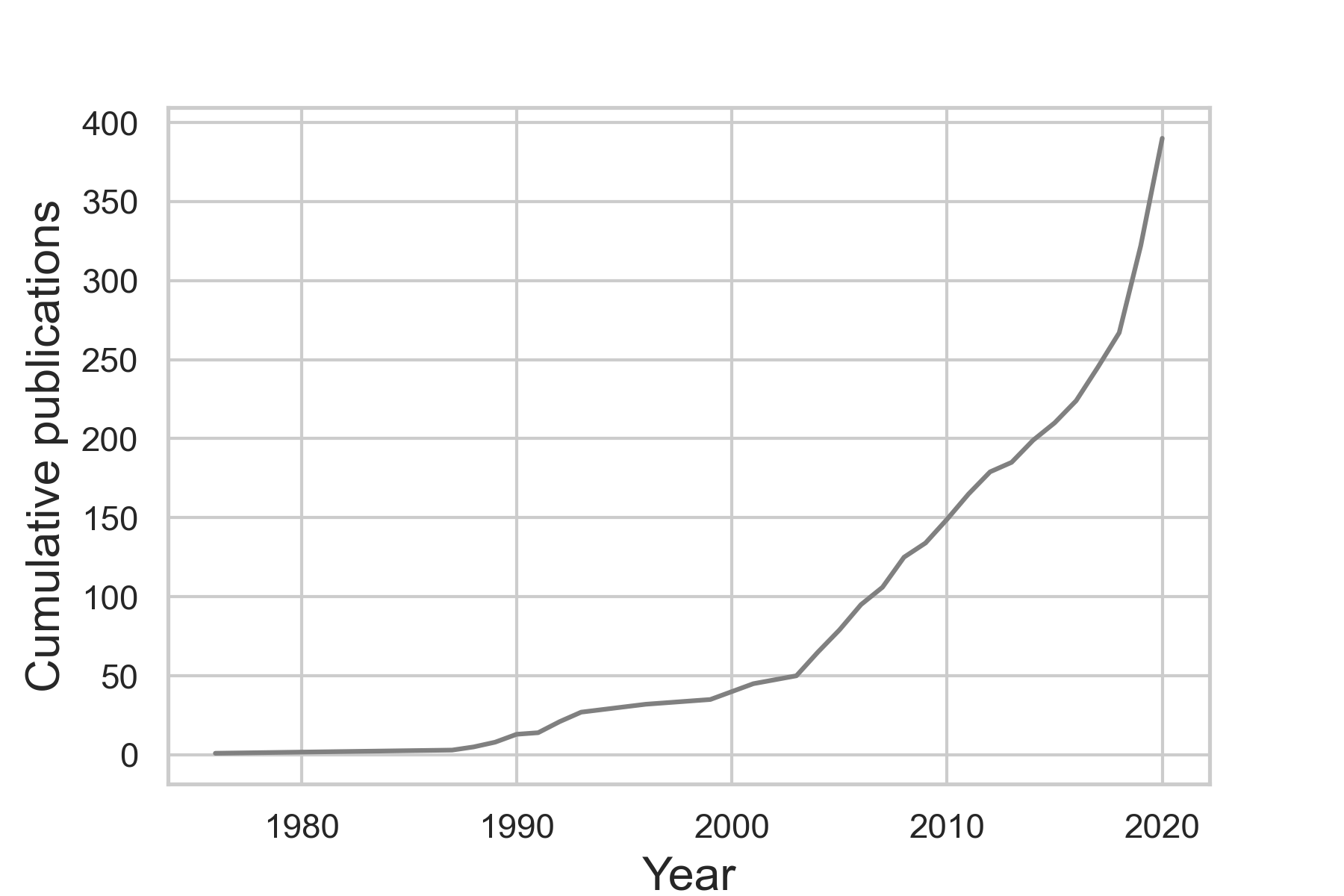}
\caption{Cumulative number of publications published per year between 1976-2020 relating to AI education (see Appendix for details about data). Note that ``AI education'' refers to the theory and practice of teaching and learning AI, and is not the same as ``AI in education'', which is about applying AI to educational settings.}
\label{fig:AIedLit}
\end{figure}

Teaching AI can be challenging, especially to learners beyond university settings, and this observation is supported by the growing body of research literature about teaching and learning of AI (see Figure \ref{fig:AIedLit}). Of course, with such a broad topic area that is also a fast changing one, designing a curriculum is not trivial. In fact, one cannot simply put forward a holistic framework onto which all AI education should be applied  \cite{langley_integrative_2019}. Rather, learners in a non-academic setting have varying capabilities and contrasting intended outcomes, and therefore we must consider the specific target audience carefully when designing an AI curriculum and course.

In this paper, we describe our experiences of developing and teaching an executive-level AI education program in the United Arab Emirates (UAE). The program was conceived by the UAE government in collaboration with the University of Oxford, UK, and forms part of the UAE's National Strategy on AI \cite{shamout_strategic_2021}. To the best of our knowledge, this paper contributes the first experience report of teaching AI in executive education, and of teaching AI in the Arab World.

\section{Course Overview}

The AI Program is a course that aims to provide an overview of AI, beginning from the basics of terminology and its history, through to the current, most commonplace, methods in ML. The learning outcomes of the course are as follows:

\begin{enumerate}
    \item To gain a broad knowledge and understanding of the core issues and concepts associated with AI;
    \item To gain detailed knowledge and a critical understanding of the current possibilities and limits of what is possible within the field of AI;
    \item To develop an appreciation of the ethical dilemmas and the challenges around privacy and cybersecurity as applied to the field of AI;
    \item To develop an appreciation of the challenges for government and governance in implementing successful technology solutions based on AI;
    \item To learn from case studies and examples of AI solutions both successful and unsuccessful; and
    \item To be empowered to apply the knowledge gained from the program in a work environment.
\end{enumerate}

This final learning outcome is a key feature of the AI Program, as we incorporate an active learning approach \cite{bonwell_active_1991} with project-based learning \cite{thomas_review_2000} where the final outcome of completion of the program is to deliver a novel project proposal for applying AI and ML.

\subsection{Course Structure}

The course was designed to be scheduled over 8 full-time teaching weeks, where each week is held in consecutive months so that students can stay engaged with their employment while participating in the course. The first 7 teaching weeks are held on location in Dubai, UAE, while the final teaching week is organized as a visit by the student cohort to Oxford, UK. Each teaching week has a broad theme, with classes spanning 5 consecutive days timetabled from 0900 to 1730. Each week's classes generally follow a standard time-boxed structure that we call the ``shape of the day'' (see Table \ref{tab:schedule} for an example schedule). Each time box consists of a plenary lecture, group work (practical exercises or for working on assignments in teams), or tutorials where students can, individually or in small groups, have one-on-one time with course tutors. Plenaries alternate between ``non-technical'' and ``technical'' to vary the type of the lecture content throughout the teaching week.

Enrollment on the AI Program did not require any prerequisite knowledge of AI. Instead, students were selected based on strategic priority for the UAE government from governmental agencies and corporations. Students were typically management-level workers with at least an undergraduate degree. No single textbook is used as reference for the teaching program; however, we provide a recommended reading list that includes machine learning (ML) textbooks such as ``Pattern Recognition and Machine Learning'' by \citeauthor{bishop_pattern_2006} (\citeyear{bishop_pattern_2006}), and ``Deep Learning'' by \citeauthor{goodfellow_deep_2016} (\citeyear{goodfellow_deep_2016}). Team assignments are set between teaching weeks to scaffold the bridge between topics, and each of which is also intended to make progress towards completing the students' final assessment, a capstone project.

\begin{table}[t]
\begin{tabular}{|c|c|c|c|c|c}
\hline
\textbf{Time/Day} & \textbf{Sun.} & \textbf{Mon.} & \textbf{Tues.} & \textbf{Wed.} & \multicolumn{1}{c|}{\textbf{Thu.}} \\ \hline
0900-0915     & Welcome         & U1              & U2               & U3                 & \multicolumn{1}{c|}{U4}                \\ \hline
0915-1045     & P1              & P2              & P3               & P4                 & \multicolumn{1}{c|}{P5}                \\ \hline
1100-1230     & TP1             & TP2             & TP3              & TP4                & \multicolumn{1}{c|}{TP5}               \\ \hline
1330-1530     & G1              & G2              & G3               & G4                 & \multicolumn{1}{c|}{T5}                \\ \hline
1530-1730     & T1              & T2              & T3               & T4                 &                                        \\ \cline{1-5}
\end{tabular}
\caption{An example of the ``shape of the day'' throughout a typical teaching week. Topical updates – U\emph{n}, Plenary lectures – P\emph{n}, Technical plenary - TP\emph{n}, Group exercises – G\emph{n}, Small-group tutorials – T\emph{n}, where \emph{n} indicates session number. Note that the working week in Muslim-majority countries is typically from Sunday to Thursday.}
\label{tab:schedule}
\end{table}

The AI Program has been running for three years (in 2018-19, 2019-20, and 2021), with the most recent teaching year having taken place during the COVID-19 pandemic and consequentially moved to an online format. As such, the curriculum was revised year-on-year to account for the associated logistical challenges, as well as to respond to student and tutor feedback. 

While the course aims to provide a broad introduction to AI, we also aim to give students enough knowledge to be able to apply their learning to their work. With this in mind, we chose to focus on ML, since this sub-field within AI is the most widely applied. The course was designed to take place over 8 teaching weeks spread out over 8 months. Each teaching week had specific themes, which were augmented with cross-cutting themes present throughout. Students are split into assigned teams to work on together throughout the course, where they are assessed by a team capstone project.

\subsection{Course Outline}

Each year of the AI Program broadly covered the same themes, with minor re-ordering of some of the curriculum, and we present our idealized full-length (8 teaching weeks) curriculum outline here.

In Week 1, we cover high-level introductions to AI subject matter, defining some key terms in the field, and laying down the theoretical and practical frameworks that the students use to engage with AI in the course. Topics during this teaching week include general introductions to AI in public discourse, society, a historical overview, and how to keep up with developments in the fields of AI and ML. Contrasts are made between ML and other areas of AI that focus on probability, rules, and search. The technical plenaries focus on introducing the different types of ML by presentations of working examples. Group work focuses primarily on team building, since it's the first-time students will have been introduced to their capstone teams. BPMN – Business Process Modelling Notation \cite{chinosi_bpmn_2012} – is introduced as the framework for process modelling and ideation around finding opportunities for AI.

Week 2 focuses on data science, examining the fundamentals of the data science process, from data collection, pre-processing, statistical analysis of data. We also look at good data governance, including anonymizing data, and the ethics of big data. This week lays the foundations to getting hands-on with data and understanding its utmost importance for ML tasks that are commonly part of data science pipelines.

By Week 3, our students will have had a broad introduction to AI/ML and associated issues around data. In this teaching week, we start delving into the technical aspects of ML – higher dimensionality and operations on data matrices, regression, classification, and how to fit models to data. 

Weeks 4 and 5 go into even more detail on supervised learning, and unsupervised and reinforcement learning respectively, while at the same time exploring issues around building ML software such as scaling up processing on the cloud and testing ML-based systems.

In Week 6, students learn the skills and key considerations necessary for successful implementation of AI projects. They learn to identify projects and opportunities that would benefit from AI, either as a project component or as a holistic solution to a problem. Students gain insights into both successful and unsuccessful AI projects through case studies, with a view to becoming effective at evaluating AI project implementation. Week 6 concludes the taught part of the course and includes a retrospective look at the course content.

Finally, the end of the course consists of the students presenting the finalized capstone projects in Week 7, and in Week 8 we host a visit to the University of Oxford, where students receive lectures from local academics and are presented with a certificate of completion. 

Throughout all the teaching weeks we also had 4 cross-cutting themes:

\begin{enumerate}
    \item Ethical, social, and legal implications of AI; 
    \item Technical tools for implementing AI, namely around using RapidMiner and Python;
    \item Business process modelling and BPMN; and
    \item Industry case studies.
\end{enumerate}

\subsection{Plenary Lectures}
\begin{figure}[t]
\includegraphics[width=\columnwidth]{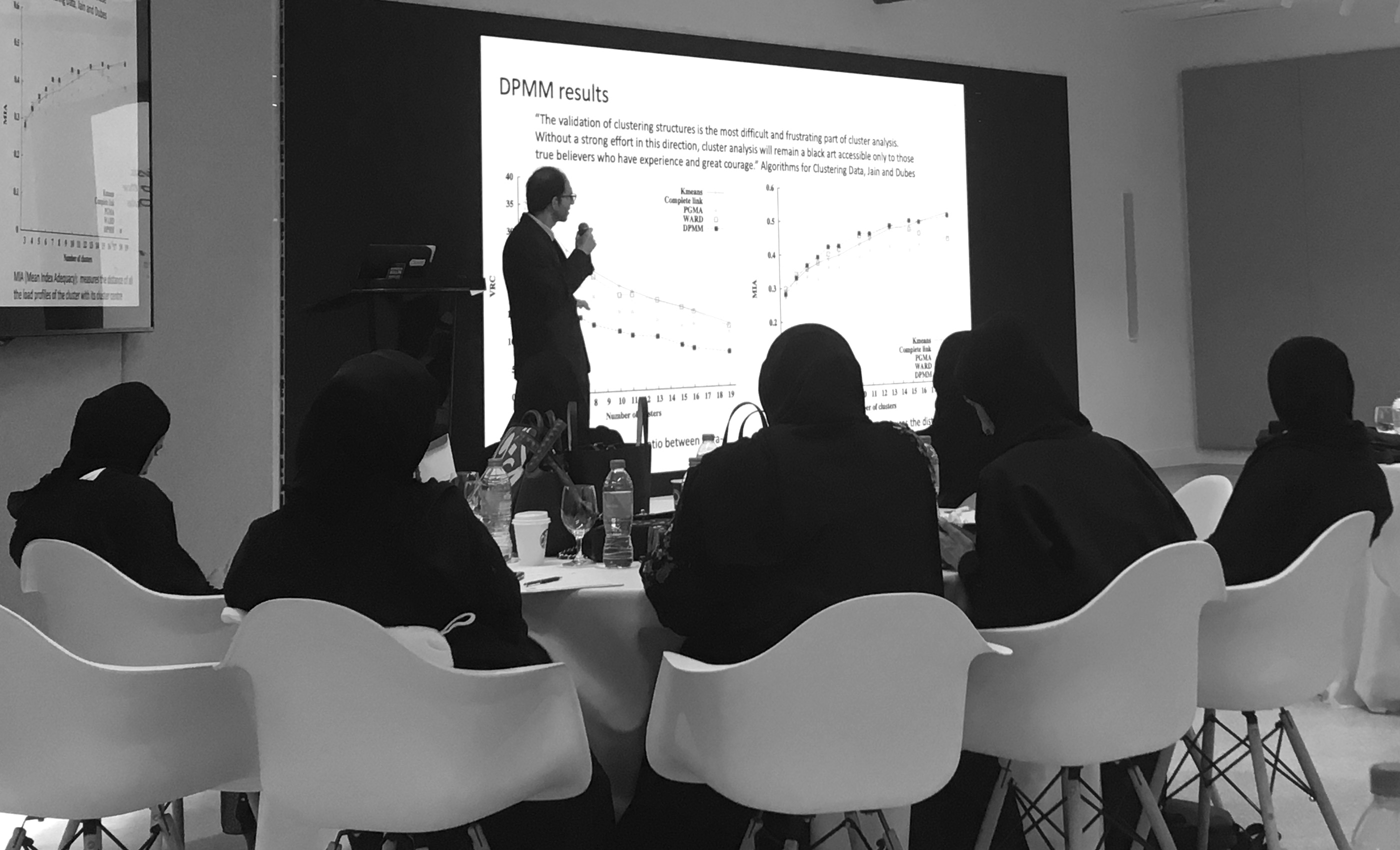}
\caption{A course tutor delivering a lecture on unsupervised learning for electricity analytics and showing a clustering example based on work by  University of Oxford researchers, in Dubai, UAE, January 2020.}
\label{fig:lecture}
\end{figure}

During each teaching week, lectures were generally delivered in the mornings within 90-minute time boxes. Each lecture consists of presentations, discussions and thought exercises. As part of the shape of the day, in the mornings we split up the more difficult technical content for our students by having one non-technical plenary (marked as P\emph{n} in Table \ref{tab:schedule}) followed by a technical plenary (marked as TP\emph{n} in Table \ref{tab:schedule}) on any given day. The theory presented in these morning sessions is then reinforced by practical group work in the afternoons. For example, a morning lecture on clustering, such as that shown in Figure \ref{fig:lecture}, would then be supported by an afternoon practical exercise where the students work with a real dataset and tools to try out some of the methods taught in the morning.

% AS: do we need authorisation to use this pic? Maybe use a pic from one of the authors instead? Also any issues around using pictures of students (even though they’re not recognisable)?

\subsection{Topical Updates}
To kick off the opening plenary lectures, we often included short sessions that focused on a topical article relating to AI to both update students on the latest developments or discourse, but also to demonstrate that AI is a fast-moving field that also has impacts on everyday life across the world. These were delivered as short presentations with a mediated whole-class discussion. 

\subsection{Group Exercises}
To take advantage of a broad range of backgrounds of students, a social constructivist approach \cite{vygotsky_mind_1978} can be taken to facilitate learning. One of the approaches we take with teaching AI is to have students work in groups to encourage the sharing of knowledge amongst peers and to get used to generating new ideas. Innovating with AI is not done in isolation. The group exercises included team building sessions to aid with the development of capstone projects, ideation workshops, business process modelling exercises, and getting hands-on experience of AI, which is a key step to gaining a deeper understanding of the theory taught in the morning lectures. The practical ML exercises were delivered in two different working environments to try and capture the broad range of technical capability of a diverse group of students.

\subsubsection{Programming for ML.}
From the first year of the AI Program we set up a working environment in which our students could get hands-on experience with ML using Python in Jupyter Notebooks \cite{kluyver_jupyter_2016}. Teaching with Jupyter has had widespread uptake in recent years \cite{brunner_teaching_2016,holdgraf_portable_2017}, and Python has been the most preferred language for ML for a number of years \cite{raschka_machine_2020}. For example, the practical ML exercises that followed the lecture shown in Figure \ref{fig:lecture} focused on getting hands-on with 3 different clustering methods, where the the Scikit-learn Python library \cite{pedregosa_scikit-learn_2011} provides capabilities for each of k-means (\verb!sklearn.cluster.KMeans!), expectation maximization (\verb!sklearn.mixture.GaussianMixture!), and agglomerative hierarchical clustering (\verb!sklearn.cluster.AgglomerativeClustering!).

\subsubsection{ML as Visual Workflows.}

Part of our feedback after the first year of the AI Program indicated that while it was clear that being able to get some experience of programming ML is highly desirable, many students found the learning curve for ML in Python too steep and this highlighted a need to cater for students who are not comfortable with programming. To cater for these types of students in subsequent iterations of the course, we sought out a working environment in which our students could get hands-on experience with ML without doing any programming. We chose to use RapidMiner, originally called Yet Another Learning Environment - YALE - \cite{mierswa_yale_2006}, a data mining software application that allows users to build visual workflows to process data. Working with visual workflows also made it easier to jump from thinking of business processes as visual flows (with BPMN diagrams) to visualizing the processes involved with dealing with data and building machine learning models. Following the clustering exercise example above in Jupyter, the same practical exercises were developed using RapidMiner, which provides visual workflow operators for each of aforementioned clustering methods.

\subsection{Capstone Project}
The team capstone project forms the cornerstone of the program, as it acts as the primary means to student assessment and gives the students a tangible takeaway from the course. Work begins on the capstone project after teaching week 1 where, after team formation, ideation begins. A series of assignments are given to the students to complete in between teaching weeks to help make progress on their capstones. These are summarized in Table \ref{tab:assignment}.

The first assignment consists of an exercise for the students to learn how to use BPMN to diagram business processes. This is set individually as teams are still being organized after Week 1. Each subsequent assignment then contributes sections to the final project report that is submitted for assessment in Week 7, where the teams also present their finalized project proposals.

The overall task for the capstone projects is to identify, examine and report on a way to apply AI to solve a problem or to improve or remove an existing process in a business or other organization. Each team chooses a topic related to AI which is of relevance to their interests, professional experience, or workplace needs. Throughout the process, guidance and support is given by course tutors.

\begin{table}[t]
\begin{tabular}{|l|c|c|}
\hline
\textbf{Project component}  & \textbf{Weight} & \textbf{Submitted in} \\ \hline
BPMN diagram*               & 5\%             & Week 2                \\ \hline
Ideation report             & 5\%             & Week 3                \\ \hline
Project proposal            & 10\%            & Week 4                \\ \hline
Project BPMN diagrams       & 5\%             & Week 5                \\ \hline
Draft project report        & 5\%             & Week 6                \\ \hline
Final project report        & 40\%            & Week 7                \\ \hline
Group presentation          & 20\%            & Week 7                \\ \hline
Reflective practice report* & 10\%            & Week 8                \\ \hline
\end{tabular}
\caption{A summary of the capstone project components, assessment weightings, and submission dates by teaching week. Components marked with * are individually assessed, while all others are assessed as a team deliverable.}
\label{tab:assignment}
\end{table}

The capstone project combines practical and academic aspects from across the program. In their teams, students develop their project on a topic related to AI and apply the resulting information to the improvement or automation of organizational processes.

Submissions for assessment include a team project report that must convey an understanding of the topic, with reference to relevant literature drawn from a range of sources. It also needs to show evidence of the effectiveness of their team and a critical evaluation of overall project management and outcomes. Specifically, the reports must: identify the opportunity to use AI; justify the use of an AI solution; explain how it links with the strategic goals of the organization; use BPMN to show business processes before and after applying the proposed solution; describe potential datasets, pre-processing considerations, and possible sources of error or bias arising; what ML models might be used and why; and finally discuss any ethical, social, legal and security implications of the proposed use of AI. During Week 7, teams make a presentation to a panel of tutors and to the whole class.

The independent reflective practice report gives students an opportunity to reflect on their personal roles in their capstone projects, and to submit a critical analysis of their individual learning. This requires them to determine what they hoped to learn from the project, and to assess how well their individual learning objectives were met. Such reflective practices have been shown to help learners revisit and challenge their preconceived ideas, examine new perspectives, and deepen their knowledge \cite{cunliffe_becoming_2004,pavlovich_developing_2009} and we believe is an important final activity in our AI course.

\subsection{Adaptation for COVID-19}
The first two teaching years of the course (2018-19 and 2019-20) took place mostly as planned, while the third year (in 2021) had to be adapted to an entirely online format because of COVID-19 restrictions limiting possibilities for tutors and students to travel. Apart from moving the course online, we also reduced the length of the course to 5 teaching weeks over 5 months, where we compressed the ML part of the course from 3 weeks into two, and removed both Week 6 on implementing AI projects and the final visit week to Oxford. However, we additionally introduced weekly ``virtual office hours'' drop-ins, which have been shown to increase student satisfaction \cite{li_does_2009}, between teaching weeks to help further motivate and support students to make progress on their capstone projects.

\section{Discussion}
Teaching on the AI Program over the past three years has given us comprehensive experiences that we wish to share for those working on delivering AI education, particularly to adult learners and to international audiences.

\subsection{Student Retrospectives}
One of the most interactive sessions that involved the whole class at the same time was a course retrospective exercise, inspired by a similar such session taught as part of an agile methods course \cite{martin_teaching_2017}, that was carried out in Week 6 of the course. This session followed the format of a project retrospective \cite{kerth_project_2013}, which is typically a consultation at the end of a project that is run by a facilitator to mediate feedback about how a project went. In our case, one of the course tutors acted as the facilitator. Groups of students were asked to think back, without the aid of their own notes, and reconstruct the shape of each teaching week during the program. 

After reconstructing each week, the group was then asked to annotate their reconstructed schedule with positive and negative comments. The facilitator then walks through each teaching week’s comments with the whole class for a general discussion.

Both the students and the teachers found this exercise extremely valuable as it gave the students a chance to reflect on the course content and discuss within their groups, and of course also acts as a mechanism for providing feedback on the course to the teachers. We recommend such retrospectives either as a program-end session, or as one that can be run at the end of each teaching week to then be able to evolve upcoming remaining parts of a teaching program on-the-fly.

\begin{figure}[h]
\includegraphics[width=\columnwidth]{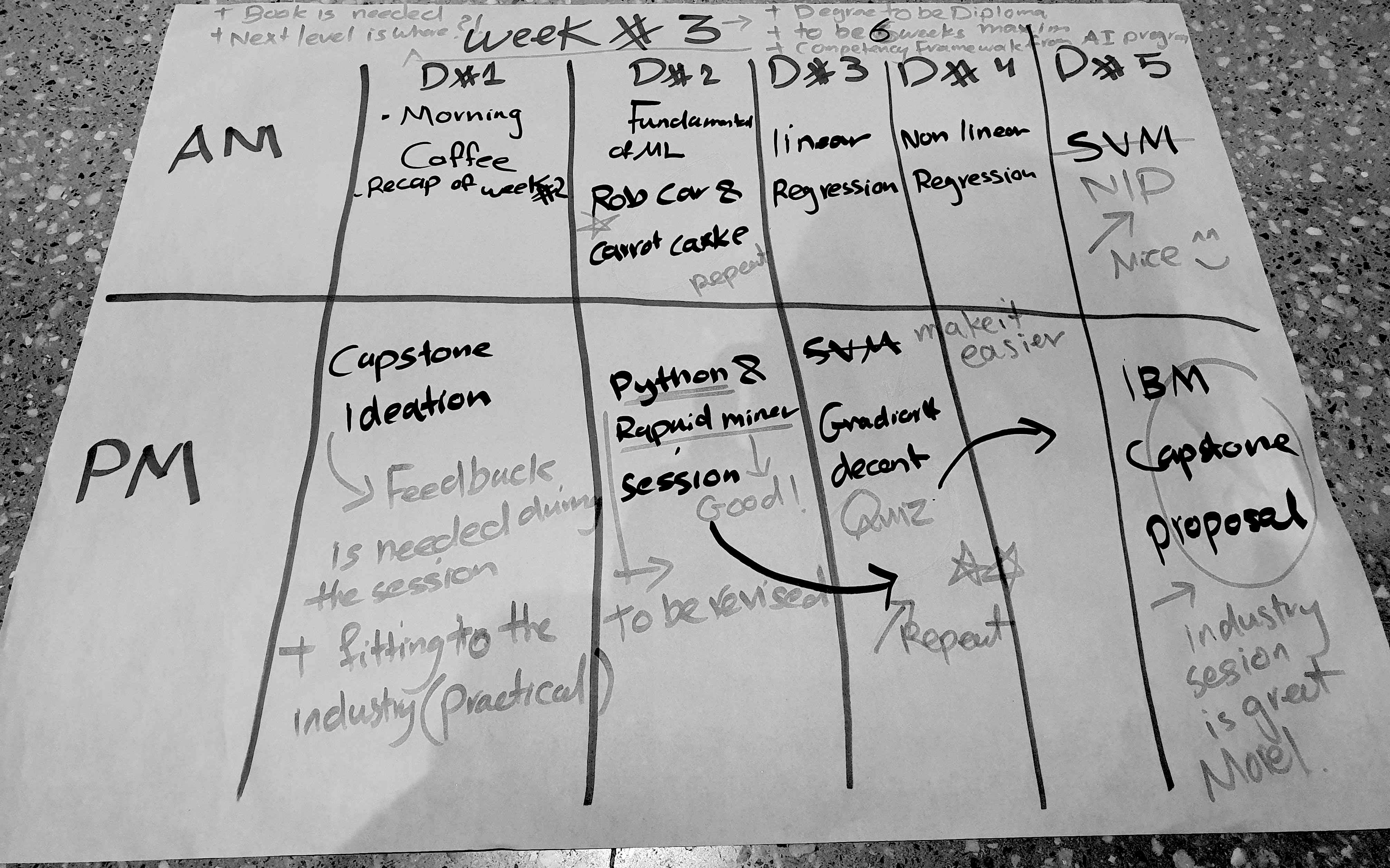}
\caption{An example of a reconstructed schedule for Week 3 of the AI Program 2019-20. A table is drawn corresponding to 5 working days and morning/evening sessions. Students then write what they remember from each part of the schedule. Finally, comments are added about what they thought of elements of the course during the teaching week.}
\label{fig:retrospective}
\end{figure}

Two main themes that arose from our retrospectives: (1) that fast and frequent feedback from course tutors is desired; and (2) that more real-world examples relevant to the region, i.e. Gulf Cooperation Council (GCC) countries - Bahrain, Kuwait, Oman, Qatar, Saudi Arabia, and the UAE, would bring further added value to the students. This aligned with our own reflections on developing and teaching the program, of which we elaborate below.

\subsection{Teaching AI to Adult Learners}
While our students were enrolled on the AI Program with no specific educational pre-requisites, all of our students are in employment, and most of whom are in management or executive positions in their respective workplaces. As such, the way in which we presented the material and exercises had to be tailored to account for the needs of adult learners.

Computational thinking is often discussed as a key skill in learning about topics relating to computing and data. Pedagogy around the concept of computation has been widely discussed \cite{lu_thinking_2009,barr_computational_2011,grover_computational_2013,selby_relationships_2015,yasar_computational_2016,patarakin_computational_2019} since \citeauthor{wing_computational_2006}'s influential \citeyear{wing_computational_2006} article on computational thinking \cite{wing_computational_2006}. However, the etymology of the word ``pedagogy'' – that is, ``ped'' meaning child and ``agogus'' meaning leader-of – implies a definition pertaining to the theory and practice of teaching children, or, rather, assumed pure novices. Pedagogical models of teaching contrast models of adult learning, where the latter acknowledge that adult learners have different, more intrinsic, motivations, bring individualized experiences, and have a more immediate need to link their learning to their everyday lives. The concept of ``andragogy'' contrasts pedagogy as being about acknowledging these needs of adult learners in teaching methods \cite{knowles_modern_1970} and is more recently discussed in terms of teaching digital skills to adults \cite{blackley_digital_2015}. We therefore put forward that these alternative needs must be considered with adult learners when teaching AI.

Through observation and from our student feedback, we find that our students' motivations closely matched those defined by andragogical theory. Our students needed to understand the importance of concepts prior to learning. They felt responsible in their learning and thus wanted to focus their time on what was important to them individually. Their prior experiences were of prime importance, where they held high expectations that what they would learn could be applied immediately to their own contexts. For example, some of our students worked for the local government-controlled Dubai Electricity and Water Authority (DEWA), a public infrastructure organization, and Abu Dhabi National Energy Company (TAQA), an energy company. With a particular interest in the energy industry, these students particularly appreciated the application domain example used as the focus of the clustering lecture shown in Figure \ref{fig:lecture}.

We believe that we successfully support adult learners' needs, particularly by focusing the AI Program on a tangible outcome – the capstone project – from the course's outset beginning in Week 1, that had regular opportunities for immediate feedback throughout the program as the capstone teams developed their projects.

A key distinction that we would like to highlight between our executive-level course and many university courses, for example, is in the student's outcomes. In university AI courses, the aim is for students to achieve a high-level of technical competency so that when graduating such courses, they can go on to be employed in technical roles such as data scientists, ML engineers and software developers, or to become AI/ML research scientists. For executives, there is less focus on achieving technical competency, and more focus on strategic competencies such as having a broad understanding, understanding business innovation potential, and wider implications of AI. Through the capstone project, we provide the scaffolding to gaining these strategic competencies.

\subsection{Teaching AI for Regional Contexts}
The AI Program was conceived through a partnership between the UAE and the University of Oxford, where the team from the university designed and delivered the teaching program to students in the UAE. A higher layer of consideration above the individualized needs of our many adult learners is that, as a group, a homogeneous feature lies in the fact that all our students are from the same country – the UAE – and mostly working in management-level positions or more senior. This meant that it was critically important to provide examples and tasks to be highly applicable to our students being within the UAE. While our teaching team already included tutors with experience/prior knowledge of the UAE, we localized aspects of the course in other ways. 

Firstly, in each year of the program we invited guest experts, from both industry and academia based in the GCC region, to speak on topics relating to AI. We hosted speakers from local offices of IBM and SAP who reported on ongoing AI projects and initiatives within the UAE. We also had specialist talks from regional academics who presented on topics on Persian language machine translation, Arabic Natural Language Processing, and research projects on ML for healthcare, such as using AI for predicting deterioration of COVID-19 patients \cite{shamout_artificial_2021}.

We also designed the course assessment to provide ample opportunity on which student groups can pose their own business problems. For the most part, this meant that students were always working towards proposing an AI solution that was specifically applicable to a local organization, or local/regional issue, within the UAE. Examples include proposals to develop document classifiers to automate categorization of police reports; to use deep learning to annotate geological images for hydrocarbon exploration; and even to develop a system to assist in the management of childhood nutrition.

\subsection{Teaching AI at Multiple Ability Levels}
While having discussed the homogeneity of the class in terms of country of origin, in practice classes our cohorts of students are highly heterogeneous when we consider background education and work experiences. While there is a minimum expectation of having a university degree, and to be in work in a UAE-based organization or governmental department, the subject matter of the background degrees held, and the nature of individual's work, varies. We therefore found that the pre-existing technical ability, as well as the ability to learn technical concepts, varied hugely. Some students had computing or engineering backgrounds and were already familiar with computing concepts, while others had business or domain focused backgrounds. As such, the latter type of students struggled when presented with ML concepts presented as mathematics or as programming code.

To account for this, and as alluded to earlier, we present AI concepts on multiple levels of abstraction to account for multiple levels of technical ability – a differentiated instruction approach. Differentiated instruction is an approach to teaching that acknowledges heterogeneous classes and applies an approach that gives students multiple options to gaining an understanding of learning material \cite{subban_differentiated_2006}. In our case, we present practical ML exercises as both visual workflows in RapidMiner and as Python code in Jupyter. We found that the use of RapidMiner enabled the less technically savvy students to experience building hands-on ML pipelines without coding, while those more able could dive into deeper details with the concomitant code examples. Likewise, having two streams of plenary lectures (``non-technical'' and ``technical'') enabled students who wanted a deeper understanding of AI methods the chance to get that detail, while those who only sought out a broader understanding of AI and related issues could benefit from the non-technical lectures.

\section{Conclusions}
As AI continues to dominate our everyday lives, more and more people will strive to understand it. AI education is now happening at the full range of education levels, from pre-school children through to adults who are in established career paths. In this paper, we reported three years of our experiences of teaching an executive-level education program in the UAE. Reflecting on those experiences, we highlight several important considerations when teaching AI:
\begin{itemize}
    \item Taking special consideration for adult learners;
    \item Being mindful of regional contexts,  in our case in the GCC locality; and
    \item Supporting highly diverse student cohorts.
\end{itemize}

We hope that others involved in designing similar such courses will learn from our experiences and promote highly effective new AI education programs.

\section{Appendix: Availability of Supporting Data}

Data about AI education-related publications up to the end of 2020 was compiled by performing an aggregated search on 6th September 2021 of bibliographic records indexed by Scopus, Web of Science, Educational Resources Information Center (ERIC), ACM Digital Library, and IEEE Xplore. This data used to plot the chart used in Figure \ref{fig:AIedLit} of this manuscript is available in Zenodo under the CC BY 4.0 license \cite{johnson_david_2021_5777468}.

\section{Acknowledgments}
The authors would like to thank the UAE Minister of State for AI in the UAE, his excellency Omar Sultan Al Olama for his support for the program, and funding through the UAE National Program for Artificial Intelligence and the Dubai Future Foundation. We would also like to thank Professor Jonathan Michie for supporting the development of the teaching program by Oxford. Thanks to Ana Grouverman, Omar Makhlouf, Angad Singh, Jude Fletcher, Ramon Granell, Sepi Chakaveh, Hammou Messatfa, Inaana Abboud, Simon O'Doherty, and Professor Nizar Habash, for their contributions as speakers/tutors. Finally, thanks to Rob Collins and Professor Cezar Ionescu for their contributions as both speakers and in developing the AI Program.

\bibliography{aaai22}

\end{document}